\documentclass[aps,prl,twocolumn,psfig,groupedaddress]{revtex4}
\def\bc{\begin{center}}
\def\ec{\end{center}}
\def\be{\begin{equation}}
\def\ee{\end{equation}}
\usepackage{longtable}
\usepackage{psfig}

\begin{document}

\title{Fractional statistics in the fractional quantum Hall effect}
\author{Gun Sang Jeon, Kenneth L. Graham, and Jainendra K. Jain}
\affiliation{Physics Department, 104 Davey Laboratory, The Pennsylvania State University,
University Park, Pennsylvania 16802}


\date{\today}

\begin{abstract}
A microscopic confirmation of the fractional statistics 
of the {\em quasiparticles} in the fractional quantum Hall effect has so 
far been lacking.  We calculate the statistics of the composite-fermion 
quasiparticles at $\nu=1/3$ and $\nu=2/5$ by evaluating the Berry phase for a 
closed loop encircling another composite-fermion quasiparticle. 
A careful consideration of subtle perturbations in the trajectory 
due to the presence of an additional quasiparticle is crucial for 
obtaining the correct value of the statistics.  The conditions for the 
applicability of the fractional statistics concept are discussed.
\end{abstract}
\pacs{71.10.Pm,73.43.-f}


\maketitle


The fractional statistics concept of Leinaas and Myrheim\cite{Leinaas} 
relies on the property that when particles with infinitely strong short range  
repulsion are confined in two dimensions, paths with different winding numbers are 
topologically distinct and cannot be deformed into one another.  
The particles are said to have statistics $\theta$ if a path 
independent phase $2\pi\theta$ results 
when one particle goes around another in a complete loop.
A half loop is equivalent to an exchange of particles, assuming 
translational invariance, which produces a phase factor $e^{i\pi \theta}=(-1)^{\theta}$.
Non-integral values of $\theta$ imply fractional statistics.
There are no fundamental particles in nature that obey fractional statistics.
Any fractional statistics objects will have to be emergent collective 
particles of a non-trivial condensed matter state.  Furthermore, they will 
be necessarily confined to two dimensions: in higher dimensions the  
notion of a particle going around another is topologically 
ill defined, because any loop can be shrunk to zero without 
ever crossing another particle.

Even though the explanation of the fractional quantum Hall effect\cite{Tsui} (FQHE)
and numerous other remarkable phenomena 
follows from the composite fermion theory with no mention of fractional 
statistics \cite{Heinonen}, fractional statistics is believed to be one of the
consequences of incompressibility at a fractional filling \cite{Arovas,Halperin,Su},
and may possibly be observable in an experiment specifically designed for this
purpose.  For Laughlin's quasiholes\cite{Laughlin} at 
$\nu=1/m$, $m$ odd, the statistics was derived explicitly by Arovas, Schrieffer, and 
Wilczek\cite{Arovas} in a Berry phase calculation, but a similar demonstration of  
fractional statistics has been lacking at other fractions, or even for   
the quasiparticles at $\nu=1/m$.
The need for a microscopic confirmation was underscored by
Kj{\o}nsberg and Myrheim\cite{Kjonsberg1} 
who showed that, with Laughlin's wave function, the quasiparticles at $\nu=1/m$
do {\em not} possess well-defined statistics.  
The reason for the discrepancy remains unclear, but it illustrates that
the fractional statistics is rather fragile and cannot be taken for granted.

The objective of this article is to revisit the issue armed with 
the microscopic composite-fermion (CF) theory of the FQHE \cite{Jain}.
A step in that direction has been taken by Kj{\o}nsberg and 
Leinaas\cite{Kjonsberg2}, whose calculation of the statistics of 
the ``unprojected" CF quasiparticle of $\nu=1/m$, the wave function for which is 
different from that of Laughlin's, produced a definite value,  
the sign of which, however, was inconsistent with general considerations.  
We confirm below that the statistics is robust to projection into the lowest 
Landau level (LL), and provide a non-trivial resolution to the sign 
enigma, which has its origin in very small perturbations in the 
trajectory due to the insertion of an additional CF quasiparticle.
The calculation is extended to $\nu=2/5$ for further verification of 
the generality of the concept.

Because the CF theory provides an accurate account of the low energy 
physics, including incompressibility at certain fractional fillings, 
it must also contain the physics of fractional statistics, which indeed 
is the case.  The fractional statistics 
can be derived heuristically in the CF theory as follows \cite{Goldhaber}.
Composite fermions are bound states of electrons and an even number ($2p$)
of vortices.  When a composite fermion goes around a closed path 
encircling an area $A$, the total phase associated with this path is given by
\begin{equation}
\Phi^*=-2\pi (BA/\phi_0- 2p N_{enc}) \;,
\label{Phi*}
\end{equation}
where $N_{enc}$ is the number of composite fermions inside the 
loop and $\phi_0=hc/e$ is called the flux quantum.  
The first term on the right hand side is the usual Aharonov-Bohm 
phase for a particle of charge $-e$ going around in a counterclockwise loop.
The second term is the contribution from the vortices bound to composite fermions, 
indicating that each enclosed 
composite fermion effectively reduces the flux by $2p$ flux quanta. 
(A note on convention: We will take the magnetic field in the $+z$ direction, 
the electron charge to be $-e$, and consider the counterclockwise direction for 
the traversal of trajectories.)

Eq.~(\ref{Phi*}) summarizes the origin of the FQHE.
The phase in Eq.~(\ref{Phi*}) is interpreted as the 
Aharonov-Bohm phase from an effective magnetic field:
$\Phi^* \equiv -2\pi B^*A/\phi_0$.
Replacing $N_{enc}$ by its expectation value $\langle N_{enc}\rangle=\rho A$, 
where $\rho$ is the two-dimensional density of electrons, we get 
\be
B^*=B-2p\phi_0\rho\;.
\label{B*}
\ee
The integral quantum Hall effect\cite{Klitzing} (IQHE) of composite fermions
at CF filling $\nu^*=n$ produces the FQHE of electrons at 
$\nu=n/(2pn+1)$.  At these special filling factors, the effective magnetic 
field is $B^*=B/(2pn+1)$.

The fractional statistics is also an immediate corollary of Eq.~(\ref{Phi*}).
Let us consider the state with CF filling $n<\nu^*<n+1$ and denote by 
$\eta_\alpha=x_\alpha-iy_\alpha$ the positions where the composite 
fermions in the topmost 
partially filled CF level are localized in suitable wave packets.  
One may imagine a density lump centered at each $\eta_\alpha$. 
An ``effective" description in terms of $\eta_\alpha$, which will 
be called CF quasiparticles (CFQP's), can in principle be obtained by 
integrating out $z_j=x_j-iy_j$.  We can conjecture the winding properties 
of the CFQP's from the underlying CF theory as follows.  Consider 
two CFQP's, sufficiently far from one another that the overlap 
between them is negligible.  According to Eq.~(\ref{Phi*}) the phase a CFQP acquires 
for a closed loop depends on whether the loop encloses the other CFQP or not.
When it does not, the phase is $\Phi^*=-2\pi eB^*A/hc$. 
The change in the phase due to the presence of the enclosed CFQP is
\be
\Delta\Phi^*=2\pi 2p \Delta \langle N_{enc}\rangle=2\pi \frac{2p}{2pn+1} 
\ee
because a CFQP has an excess of  
$1/(2pn+1)$ electrons associated with it relative to the uniform state
[producing a local charge of $q^*=-e/(2pn+1)$].  
With $\Delta\Phi^*=2\pi\theta^*$ we get the CFQP statistics parameter  
\be
\theta^*=\frac{2p}{2pn+1}\;.
\label{theta*}
\ee
This value is consistent, $mod$ 1, with those quoted previously\cite{Halperin,Su}.

Our goal is to confirm Eq.~(\ref{theta*}) in a microscopic calculation
of the Berry phases.  The statistics is given by
\be
\theta^* =
\oint_{\cal C} \frac{d\theta}{2\pi} \frac{\left<\Psi^{\eta,\eta'}| i\frac{d}{d\theta}
\Psi^{\eta,\eta'}\right>} {\left<\Psi^{\eta,\eta'}|\Psi^{\eta,\eta'}\right>}-
\oint_{\cal C} \frac{d\theta}{2\pi}
\frac{\left<\Psi^\eta|
i\frac{d}{d\theta} \Psi^{\eta}\right>}{
\left<\Psi^\eta|\Psi^{\eta}\right>}\;,
\label{Berry}
\ee
where $\Psi^\eta$ is the wave function containing a single CFQP at $\eta$, and 
$\Psi^{\eta,\eta'}$ has two CFQP's at $\eta$ and $\eta'$.
Here we take $\eta=R e^{-i\theta}$, and ${\cal C}$ refers to the path 
with $R$ fixed and $\theta$ varying from $0$ to $2\pi$ in the counterclockwise direction.
For convenience, we will take $\eta'=0$.

The calculation of $\theta^*$ requires microscopic wave functions
which are constructed as follows. 
The composite fermion theory maps the problem of interacting electrons at 
$\nu$ into that of weakly interacting composite fermions at 
$\nu^*$.  In order to put these composite fermions at $\eta_\alpha$, we first 
construct the electronic wave function at $\nu^*$ with the 
electrons in the partially filled level at $\eta_\alpha$; these are 
placed in the coherent state wave packets 
\be
\bar \phi^{(n)}_{\eta}(\vec{r})= \phi^{(n)}_{\eta}(\vec{r})
\exp[-|z|^2/4l^{*2}]
\ee
\be
\phi^{(n)}_{\eta}(\vec{r})= (\bar{z}-\bar{\eta})^n
\exp[\bar{\eta}z/2l^{*2} -|\eta|^2/4l^{*2}]
\ee
where $l=\sqrt{\hbar c/eB}$ and $l^*=(2pn+1)^{1/2}l$ are the magnetic lengths at $B$ and 
$B^*$.  We then make a mapping into composite fermions {\em in a manner that preserves 
distances} (to zeroth order) by multiplying by $\Phi_1^{2p}=
\prod_{j<k=1}^N(z_j-z_k)^{2p}\exp[-2p\sum_i|z_i|^2/4l_1^2]$
with $l_1^2=\hbar c/eB_1=\hbar c/e \rho\phi_0$, followed by projection into 
the lowest LL.

To give an explicit example, consider two CFQP's at $\nu=1/(2p+1)$.
The electron wave function at $\nu^*=1$ with fully occupied lowest LL  
and two additional electrons in the second LL at $\eta$ and $\eta'$ is  
\be
\Phi_1^{\eta,\eta'}=\left|\begin{array}{ccccc}
\phi^{(1)}_{\eta}(\vec{r}_1) & \phi^{(1)}_{\eta}(\vec{r}_2) & . &.&.\\
\phi^{(1)}_{\eta'}(\vec{r}_1) & \phi^{(1)}_{\eta'}(\vec{r}_2) & . &.&.\\
1 & 1 & . &.&. \\
z_{1}&z_{2}&. &.&.\\
.&.&.&.&.\\
.&.&.&.&.  \\
z_1^{N-3} & z_2^{N-3} & . &.&.
\end{array}
\right|e^{-\sum_j|z_j|^2/4l^{*2}} \;.
\ee
This leads to the (unnormalized) wave function for two CFQP's at $\nu=1/(2p+1)$:
\begin{eqnarray}
\Psi^{\eta,\eta'}_{1/(2p+1)} &=& {\cal P} \prod_{i<k=1}^N(z_i-z_k)^{2p} 
\left|\begin{array}{ccccc}
\phi^{(1)}_{\eta}(\vec{r}_1) & \phi^{(1)}_{\eta}(\vec{r}_2) & . &.&.\\
\phi^{(1)}_{\eta'}(\vec{r}_1) & \phi^{(1)}_{\eta'}(\vec{r}_2) & . &.&.\\
1 & 1 & . &.&. \\
z_{1}&z_{2}&. &.&.\\
.&.&.&.&.\\
.&.&.&.&. \\
z_1^{N-3} & z_2^{N-3} & . &.&. 
\end{array}
\right| \nonumber \\
& & \;\;\;\;\times e^{-\sum_j|z_j|^2/4l^2}\;.
\label{2CFQP}
\end{eqnarray}
Here, ${\cal P}$ is the lowest Landau level projection operator, 
and we have used $l^{*-2}+2pl_1^{-2}=l^{-2}$ which is equivalent to Eq.~(\ref{B*}). 
Wave functions for one or many CFQP's at arbitrary filling 
factors can be written similarly.  The lowest LL projection can be performed 
in either one of two ways described in the literature\cite{JK}.  
Our wave functions are similar to those considered 
in Ref.~\onlinecite{Kjonsberg2}, but not identical.

The integrands in Eq.~(\ref{Berry}) involve $2N$ dimensional integrals over 
the CF coordinates, which we evaluate by Monte Carlo method.  To determine the 
O(1) difference between two O($N$) quantities on the right hand side
with sufficient accuracy,
we use the same importance sampling for both the quantities on the 
right hand side, which reduces statistical fluctuations in the difference. 
The two-CFQP wave function $\Psi^{\eta,\eta'}$ is used as the weight function
for both terms in Eq.~(\ref{Berry}).  Approximately $4 \times 10^{8}$ 
iterations are performed for each point.  For $\nu=1/3$ we have studied systems with
$N=50$, 100, and 200 particles, and the projected wave function is used.
In this case, a study of fairly large systems is possible because 
no explicit evaluation of the determinant is required at each step. 
For $\nu=2/5$, it is much more costly to work with the projected wave 
function, and we have studied only the unprojected wave function 
for $N=50$ and 100.  The calculation at $\nu=1/3$ explicitly demonstrates 
that $\theta^*$ is independent of whether the projected or the 
unprojected wave function is used, or which projection method is used;
we assume the same is true at $\nu=2/5$. 

\begin{figure}
\centerline{\psfig{figure=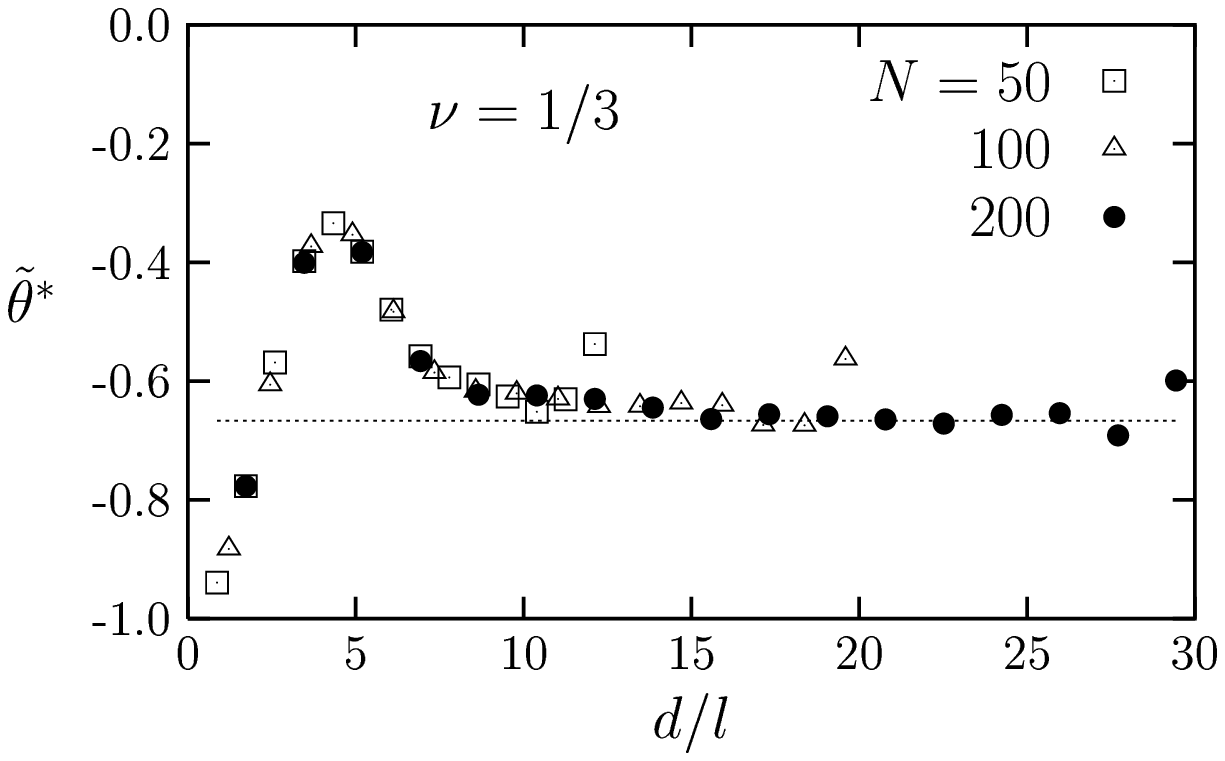,width=3.0in,angle=0}}
\vspace{-1.0cm}
\centerline{\psfig{figure=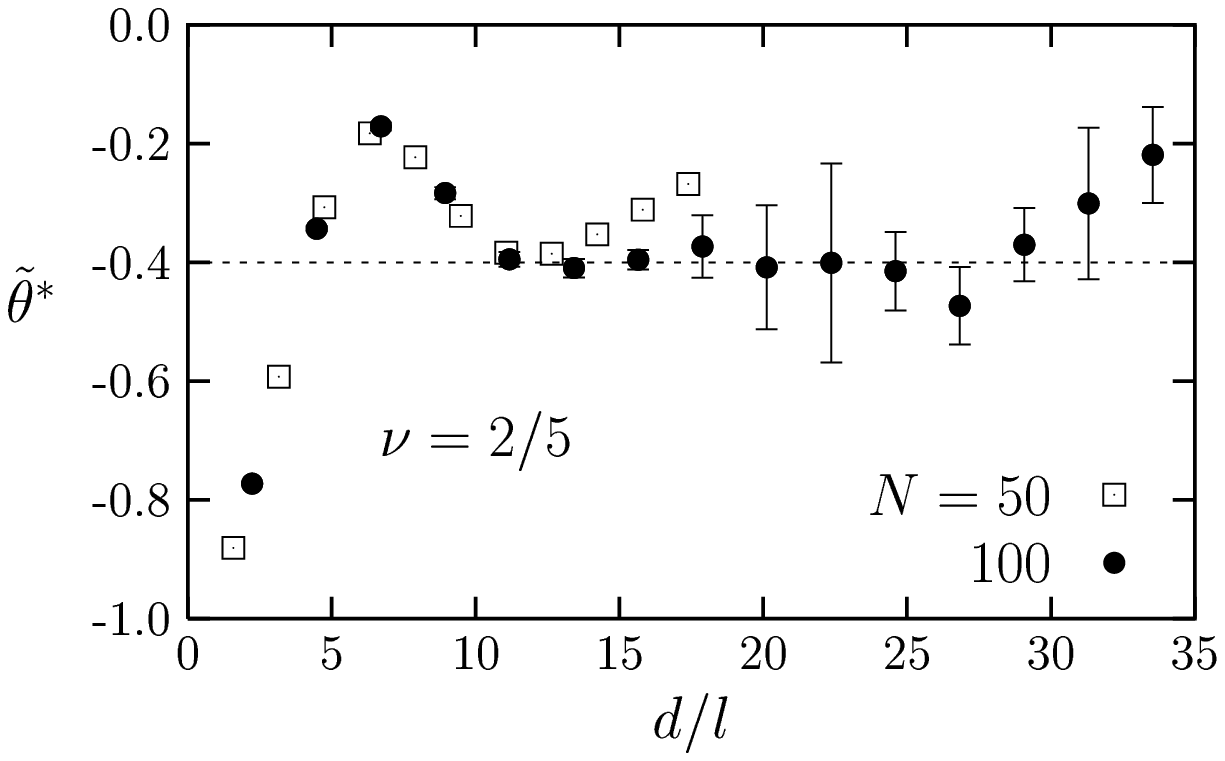,width=3.0in,angle=0}}
\caption{The statistical angle $\tilde\theta^*$ for the CF quasiparticles at 
$\nu=1/3$ (upper panel) and $\nu=2/5$ (lower panel) as a function of 
$d \equiv |\eta-\eta'|$.  $N$ is the total number of composite fermions, and $l$ is the 
magnetic length. (The symbol $\tilde\theta^*$ is used rather than $\theta^*$ 
for the statistical angle to remind that the correct 
interpretation of the results gives $\theta^*=-\tilde\theta^*$.) The error bar from 
Monte Carlo sampling is not shown explicitly when it is smaller than the symbol size.
The deviation at the largest $d/l$ for each $N$ is due to proximity to the edge.} 
\label{fig1}
\end{figure}

The statistics parameter $\theta^*$ is  shown in Fig.~\ref{fig1} for 
$\nu=1/3$ and $\nu=2/5$.  $\theta^*$ takes a well-defined value for large
separations.  
At $\nu=1/3$ it approaches the asymptotic value of $\theta^*=-2/3$,
which is consistent with that obtained in Ref.~\onlinecite{Kjonsberg2}
without lowest LL projection.  At $\nu=2/5$ the system size is 
smaller and the statistical uncertainty bigger, but the asymptotic value 
is clearly seen to be $\theta^*=-2/5$.  At short separations there are 
substantial deviations in $\theta^*$; it reaches the asymptotic 
value only after the the two CFQP's are separated by more than  
$\sim$ 10 magnetic lengths.

The microscopic value of $\theta^*$ obtained above has the same magnitude as 
$\theta^*$ in Eq.~(\ref{theta*}) {\em but the opposite sign}. 
The sign discrepancy, if real, is profoundly disturbing because it cannot be 
reconciled with Eq.~(\ref{Phi*}) and would cast doubt on the 
fundamental interpretation of the CF physics in terms of an effective 
magnetic field.

\begin{figure}
\centerline{\psfig{figure=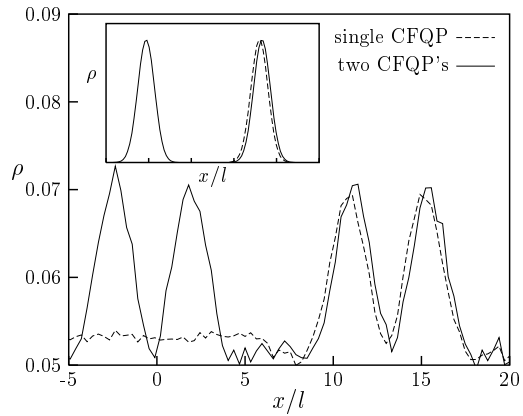,width=3.0in,angle=0}}
\vspace{-5mm}
\caption{Density profiles for $\Psi^{\eta}$ (dashed line) and 
$\Psi^{\eta,\eta'}$ (solid line) along the $x$ axis at 
$\nu=1/3$, with $\eta=13 l$ and $\eta'=0$. 
(The uniform state has density $\rho=\nu/2\pi$.)  
The noise on the curves is a measure of the statistical uncertainty in the 
Monte Carlo simulation. The CFQP in the second level 
has a smoke ring shape, with a minimum at its center.  The CFQP is located at 
$x=13 l$ in $\Psi^{\eta}$, but is shifted outward in $\Psi^{\eta,\eta'}$. 
The inset shows the density profiles for $\chi^{\eta}$ (dashed line) and
$\chi^{\eta,\eta'}$ (solid line), describing CFQP's in  
the lowest LL (see the text for definition).
\label{fig2}}
\end{figure}
 
To gain insight into the issue,  
consider two composite fermions in the otherwise empty lowest LL, 
for which various quantities can be obtained analytically. 
When there is only one composite fermion at $\eta=Re^{-i\theta}$, 
it is the same as an electron, with the wave function given by
\be
\chi^{\eta}=\exp[\bar{\eta}z/2 -R^2/4-|z|^2/4]\;.
\ee
For a closed loop, 
\be
\oint_{\cal C} \frac{d\theta}{2\pi}
\frac{\left<\chi^\eta|
i\frac{d}{d\theta} \chi^{\eta}\right>}{
\left<\chi^\eta|\chi^{\eta}\right>}=-\frac{R^2}{2l^2}=-\frac{\pi R^2B}{\phi_0}\;.
\ee
Two composite fermions, one at $\eta$ and the other at 
$\eta'=0$, are described by the wave function 
\be
\chi^{\eta,0}=(z_1-z_2)^{2p}(e^{\bar\eta z_1/2}-e^{\bar\eta z_2/2})
e^{-(R^2+|z_1|^2+|z_2|^2)/4}
\ee
Here, we expect $\theta^*=2p$. 
However, an explicit evaluation of the Berry phase shows, neglecting O($R^{-2}$) terms 
\be
\oint_{\cal C} \frac{d\theta}{2\pi}
\frac{\left<\chi^{\eta,0}|
i\frac{d}{d\theta} \chi^{\eta,0}\right>}{
\left<\chi^{\eta,0}|\chi^{\eta,0}\right>}=-\frac{R^2}{2l^2}-2p\;,
\label{wrong}
\ee
which gives $\theta^*=-2p$ for large $R$.  Again, it apparently has 
the ``wrong" sign.

A calculation of the density for $\chi^{\eta,0}$ shows that 
the actual position of the outer composite fermion is not $R=|\eta|$ but 
$R'$, given by
\be
R'^2/l^2=R^2/l^2+4\cdot 2p
\label{R2}
\ee
for large $R$.  This can also be seen in the inset of Fig.~(\ref{fig2}).
The correct interpretation of Eq.~(\ref{wrong}) therefore is 
\be
\oint_{\cal C} \frac{d\theta}{2\pi}
\frac{\left<\chi^{\eta,0}|
i\frac{d}{d\theta} \chi^{\eta,0}\right>}{
\left<\chi^{\eta,0}|\chi^{\eta,0}\right>}=-\frac{R'^2}{2l^2}+2p
\ee
which produces $\theta^*=2p$.  The O(1) correction to the area enclosed 
thus makes a non-vanishing correction to the statistics.
(It is noted that the CF quasiparticle at $\eta=0$ is also a little off center, 
and executes a tiny circular loop which provides another correction to the 
phase, but this contribution vanishes in the limit of large $R$.)

This exercise tells us that an implicit assumption made in the earlier analysis,
namely that the position of the outer CFQP labeled by $\eta$ remains 
unperturbed by the insertion of another CFQP, leads to an incorrect value 
for $\theta^*$.  In reality, inserting another CFQP inside the loop pushes 
the CFQP at $\eta$ very slightly outward.

To determine the correction at $\nu=n/(2pn+1)$, we note that the 
mapping into composite fermions preserves distances to zeroth order,
so Eq.~(\ref{R2}) ought to be valid also at $\nu=n/(2pn+1)$.
This is consistent with the shift seen in Fig.~\ref{fig2} for the position of the 
CFQP.  Our earlier result
\be
\oint_{\cal C} \frac{d\theta}{2\pi}
\frac{\left<\Psi^{\eta,0}|
i\frac{d}{d\theta} \Psi^{\eta,0}\right>}{
\left<\Psi^{\eta,0}|\Psi^{\eta,0}\right>}=-\frac{R^2}{2l^{*2}}-\frac{2p}{2pn+1}
\ee
ought to be rewritten, using $l^{*2}/l^2=B/B^*=2pn+1$, as  
\be
\oint_{\cal C} \frac{d\theta}{2\pi}
\frac{\left<\Psi^{\eta,0}|
i\frac{d}{d\theta} \Psi^{\eta,0}\right>}{
\left<\Psi^{\eta,0}|\Psi^{\eta,0}\right>}=-\frac{R'^2}{2l^{*2}}+\frac{2p}{2pn+1}
\ee
When the contribution from the closed path without the other CFQP, 
$-R'^2/2l^{*2}$, is subtracted out, $\theta^*$ of Eq.~(\ref{theta*}) is obtained.
The neglect of the correction in the radius of the loop introduces an error 
which just happens to be twice the negative of the correct answer.

The fractional statistics of the CFQP should not be confused with 
the fermionic statistics of composite fermions.  The wave 
functions of composite fermions 
are single-valued and antisymmetric under particle exchange;
the fermionic statistics of composite fermions has been 
firmly established through a variety of facts, including the observation 
of the Fermi sea of composite fermions, the observation of FQHE at fillings 
that correspond to the IQHE of composite fermions, and also by the fact that 
the low energy spectra in exact calculations on finite systems 
have a one-to-one correspondence with those of weakly interacting
fermions\cite{Heinonen}.  
There is no contradiction, however. After all, any fractional statistics in nature  
{\em must} arise in a theory of particles that are either fermions or bosons
when an {\em effective} description is sought in terms of certain collective 
degrees of freedom.  The fractional statistics appears in the CF theory when 
the original particles $\{z\}$ are treated in an average, mean field 
sense to formulate an effective description in terms of the 
CFQP's at $\{\eta\}$.

The fractional statistics is equivalent to the existence of an 
effective locally pure gauge vector potential, with 
no magnetic field associated with it except 
at the particle positions\cite{Arovas}.  In the present case, 
the substantial deviation of $\theta^*$ from its asymptotic
value at separations of up to $10$ magnetic lengths indicates 
a core region where the induced vector potential is not pure gauge, 
thereby imposing a limitation on
a model in which the the CFQP's are approximated by ideal,
point-like particles with well-defined fractional statistics (anyons).
Such an idealization is valid only to the extent that the relevant CFQP trajectories 
do not involve a significant overlap of CFQP's.  
Given that there does not exist a strong repulsion between the CFQP's -- 
the inter-CF interaction is very weak and often {\em attractive}\cite{Lee} -- 
such trajectories are not precluded energetically, and the 
anyon model is therefore not a justifiable approximation, except, possibly,  
for very dilute systems of CFQP's in a narrow filling 
factor range around $\nu=n/(2pn+1)$.
Any experimental attempt to measure the fractional statistics of the CFQP's 
must ensure that they remain sufficiently far apart during the 
measurement process.

Partial support of this research by the National Science Foundation under grants
no. DGE-9987589 (IGERT) and DMR-0240458 is gratefully acknowledged.
We thank Profs. A.S. Goldhaber and J.M. Leinaas for comments.

\end{document}